\pdfoutput=1
\documentclass[aps,prl,amsmath,twocolumn,10pt]{revtex4-1}
\usepackage[T1]{fontenc}
\usepackage[latin9]{inputenc}
\usepackage{color}
\usepackage{amsmath}
\usepackage{amssymb}
\usepackage{graphicx}
\usepackage{tikz}
\usepackage[unicode=true,pdfusetitle,
 bookmarks=true,bookmarksnumbered=false,bookmarksopen=false,
 breaklinks=false,pdfborder={0 0 1},backref=false,colorlinks=false]
 {hyperref}

\makeatletter
\@ifundefined{textcolor}{}
{%
 \definecolor{BLACK}{gray}{0}
 \definecolor{WHITE}{gray}{1}
 \definecolor{RED}{rgb}{1,0,0}
 \definecolor{GREEN}{rgb}{0,1,0}
 \definecolor{BLUE}{rgb}{0,0,1}
 \definecolor{CYAN}{cmyk}{1,0,0,0}
 \definecolor{MAGENTA}{cmyk}{0,1,0,0}
 \definecolor{YELLOW}{cmyk}{0,0,1,0}
}

\newcommand{\cscc } {$\chi$SC }
\newcommand{\csck} {$\chi$SC, }

\newcommand{\ie} {\emph{i.e. }}

\usepackage{braket}

\makeatother

\begin{document}

\title{Proposed Spontaneous Generation of Magnetic Fields by Curved Layers of a Chiral Superconductor}

\author{T. Kvorning$^1$, T. H. Hansson$^{1,2}$, A. Quelle$^3$ and C. \surname{Morais Smith}$^3$ }

\affiliation{$^1$Department of Physics, Stockholm University, AlbaNova University
Center, SE-106 91 Stockholm, Sweden, \\
$^2$Nordita, KTH Royal Institute of Technology and Stockholm University,
Roslagstullsbacken 23, SE-106 91 Stockholm, Sweden, \\
$^3$Institute for Theoretical Physics, Center for Extreme Matter and
Emergent Phenomena, Utrecht University,  Princetonplein 5, 3584CC Utrecht,
The Netherlands.}

\date{\today}
\begin{abstract}
We demonstrate that two-dimensional chiral superconductors on curved surfaces spontaneously develop magnetic flux. This geometric Meissner effect provides an unequivocal signature of chiral superconductivity, which could be observed in layered materials under stress. We also employ the effect to explain some puzzling questions related to the location of zero-energy Majorana modes.

\end{abstract}
\maketitle

\noindent
Although it has been known for quite a while that all (gapped) superconductors
are topologically ordered (see \emph{e.g.}  \cite{hansson04}),
the chiral ones are particularly fascinating. Most interesting are the odd-pairing
 chiral superconductors ($\chi$SCs) in two spatial dimensions
($2d$), and the layered ones in $3d$, such
as chiral $p$-wave, $f$-wave \emph{etc.. } Typically, these states
 support vortices that are non-Abelian anyons \cite{read00,greiter92,ivanov01,stern04}.

There are several candidate materials for chiral pairing and most of these are layered. Examples range from UPt$_{3}$ \cite{tou98}, Li$_{2}$Pt$_{3}$B \cite{nishiyama07} and Sr$_{2}$RuO$_{4}$ \cite{Maeno1994,Mackenzie2017} for odd pairing, to SrPtAs \cite{fischer14,nishikubo11} and
doped graphene \cite{kiesel12,black-schaffer14,nandkshore12,Fedorov2014,Ichinokura2016} for even pairing. 

Most of the experimental evidence for \cscc is by observation of 
spontaneous breaking of time-reversal invariance, but the experiments are inconclusive and
 it is essential to find an unequivocal 
signature for $\chi$SCs, similar to the Meissner effect in ordinary
SCs. Since the essence of the Meissner effect is the gap to flux
excitations, one can think of a SC as a flux insulator. Ordinary
charge insulators can be either trivial or topological, so it is natural
to ask whether the proper description of topological SCs would 
be in terms of \emph{topological flux insulators}.  

In this letter, we show that a $2d$ \cscc will spontaneously develop a magnetic flux when
put on a curved surface. Conversely, if a spontaneously generated magnetic field is observed,
the very fact that one of the two directions perpendicular to the surface is picked out clearly
shows that there are super-currents breaking chirality. We thus 
submit that the geometric Meissner, \emph{i.e.} the spontaneous magnetic field due to curvature, will be a smoking-gun signature of a layered $\chi$SC. 

To understand this effect, it is useful to recall that in addition to the Hall conductivity,
quantum Hall (QH) liquids are characterized by their response to the 
curvature of the $2d$ surface on which they reside. This
effect, which was first described by Wen and Zee \cite{wen92-3}, comes
about because an electron in a QH liquid carries a ``spin'' due to the cyclotron motion 
(often referred to as \emph{orbital spin}), and thus acquires a Berry phase
when moving on a curved surface. When completing a closed orbit on a surface
with constant Gaussian curvature $K$ and magnetic field $B$, it will
pick up a phase $\sim Area\times(eB+sK)$, where $s$ is the orbital spin. 
Since the QH liquids form at high magnetic fields, the contribution from 
curvature cannot be detected in an experiment.

In a \csck the situation is very different. The conditions for detecting the magnetic flux response to curvature is  much more
favourable. Because there is no background flux, our results show that the geometric Meissner effect 
could be detected in a bent layered \cscc using a sensitive SQUID.

After a short review of the effective response theories for charge and flux insulators,
we identify the origin of the geometric Meissner response, and use this finding to resolve 
some puzzling questions related to the location of zero-energy Majorana edge modes (Majorinos) 
and design a geometry-driven tunneling current in a weak link. Finally, we
discuss possible experiments to detect our theoretical predictions.

\noindent \textbf{Response action for $2d$ $U(1)$ insulators.}
Insulators are systems with a conserved $U(1)$ charge and
a gap to charged bulk excitations, implying that the response action
is local. In the standard case of the $U(1)$ electromagnetic gauge
symmetry related to electric-charge conservation, it is known that
insulators can be trivial or topological. The simplest trivial insulator
is just empty space, while others differ by having a more
complicated electromagnetic response, with material-dependent
parameters that can be continuously changed to those of the vacuum.
A non trivial, or topological, insulator cannot be continuously changed
into the vacuum, and the effective action typically has terms with
quantized coefficients that can change only at phase transitions related
to the closing of the energy gap. Typical examples in $2d$ 
are Chern insulators and integer QH systems. We first consider the known case of an electric insulator to exploit the analogy with flux insulators, \emph{i.e.,} superconductors, to which we then turn our attention.

\emph{i) Charge insulators.}
Here, the effective action $W[A_{\mu}]$ encodes current correlation functions and 
the response to external electromagnetic fields $A_{\mu}$, \emph{i.e.} the current expectation value is $\langle j_\mu \rangle = 2\pi\delta W/\delta A_\mu $. The QH response to a slowly varying current is encoded in the Chern-Simons (CS) term
\begin{multline}
	W[A_{\mu}]  =W_{CS}[A_{\mu}]+\dotsb\\
	=\frac{\nu e^2}{2h}\int dt d^2x\,\varepsilon^{\mu\nu\sigma}A_{\mu}\partial_{\nu}A_{\sigma}+\dotsb \label{eq:CS}
\end{multline}
which not only implies a Hall conductivity $\sigma_{H}=\nu e^2/h$, but also
relates the total charge $N_Q$ of a region $S$ to the total flux $N_\phi$ 
through it. Changing the number of flux quanta $N_\phi$ by $\delta N_\phi$ will, according to \eqref{eq:CS}, lead to a change  $\delta N_Q=\nu \delta N_{\phi}$ in the number of unit charges $N_Q$. 
If the electromagnetic field is the only long-distance effect, which is the case in a pristine QH experiment, this relation also holds for the total values,
\begin{align}
	N_{Q}=\nu N_{\phi} \ . \label{eq:fillingfr}
\end{align}
Note that
the sign of $\nu$ defines an orientation on the $2d$ surface and thus breaks chiral symmetry. 

We now turn to the main topic of this paper\textemdash the effect
of geometry, \ie how the system depends on a spatial (possibly
time-dependent) metric $g_{ij}$. It was shown in Ref. \cite{hoyos12}
that the long-wavelength part of the geometric response is captured
by the Wen-Zee term \cite{wen92-3} 
\begin{align}
W_{WZ}[A_{\mu},\omega_{\mu}]=\frac{e\kappa_{QH}}{2\pi}\int dtd^{2}x\,\varepsilon^{\mu\nu\sigma}\omega_{\mu}\partial_{\nu}A_{\sigma}\ ,\label{geoterm}
\end{align}
where $\omega_{\mu}$ (which depends on $g_{ij}$) is a potential for
the Gauss curvature $K$, \emph{viz.} $\varepsilon^{ij}\partial_{i}\omega_{j}=\sqrt{g}K$, and  $\kappa_{QH}$  defines the long wavelength charge response to the curvature. %
Just as the CS term, the Wen-Zee term specifies
an orientation given by the sign of $\kappa_{QH}$, so it can again only
be present if there is a preferred orientation. 
For closed surfaces, the Wen-Zee term gives rise to a shift in the relation \eqref{eq:fillingfr}, 
\begin{align}
N_{Q}=\nu N_{\phi}+\kappa_{QH}\chi,\label{eq:qhshift}
\end{align}
where $\chi=\int d^2 x \sqrt g K/2\pi$ is the Euler characteristic of $S$. Since $N_Q$, $N_\phi$ and $\chi$ are integers, and $\nu$ is rational, $\kappa_{QH}$ must be quantized.

\emph{ii) Flux insulators.} We now switch to the systems
of interest\textemdash the $\chi$SCs. In the spirit of
Ref. \cite{hansson04}, we will use a toy model where the
electromagnetic field is mimicked by $2d$ Maxwell theory.
In $2d$, conservation of magnetic flux (which is a consequence of Maxwell's equations)
\begin{align}
0=\partial_{\mu}j_{flux}^{\mu}\equiv\frac{1}{2}\partial_{\mu}\varepsilon^{\mu\nu\sigma}F_{\nu\sigma}\, ,
\end{align}
amounts to having a conserved $U(1)$ charge. Since a SC is a flux insulator, \ie  has a gap to flux excitations, it is natural to consider the effective action $W[b_\mu]$, where the external gauge field $b_{\mu}$ is coupled to $j_{flux}^{\mu}$, so that $\langle  j_{flux}^{\mu} \rangle = \delta W / \delta b_\mu $ is the expectation value of flux current. 
This coupling can be interpreted in two different ways, as seen by 
\begin{multline}
\int dtd^2x\,j_{flux}^{\mu}b_{\mu} =\int dtd^2x\,A_{\mu}\varepsilon^{\mu\nu\sigma}\partial_{\nu}b_{\sigma}\\
=\int dtd^{2}x\,j^{\mu}A_{\mu}
\end{multline}
which identifies $\varepsilon^{\mu\nu\sigma}\partial_{\nu}b_{\sigma}$
as the supercurrent.

Ordinary SCs have chiral symmetry and are trivial flux insulators. However, $\chi$SCs could also be topologically non-trivial and we now focus on the response to curvature given by the SC version of the Wen-Zee term
\begin{align}
W_{WZ}[b_{\mu},\omega_{\mu}] =\frac{\kappa_{C}\Phi_{0}}{2\pi}\int dtd^{2}x\,\varepsilon^{\mu\nu\sigma}\omega_{\mu}\partial_{\nu}b_{\sigma}\ ,\label{geomeissner1}
\end{align}
defined by the single parameter $\kappa_C$, which, just as $\kappa_{QH}$, has to be quantized. 

Eq. \eqref{geomeissner1} encodes the 
geometric Meissner response, which relates the total flux through a region $S$
to its total curvature $\chi$,
\begin{align}
N_{\Phi}=\kappa_{C}\chi\label{eq:kappa}.
\end{align}
$N_{\Phi}$ denote flux in units of the superconducting flux quantum $\Phi_0 = h/2e$. Changing the sign of $\kappa_C$ defines the direction of the magnetic field and it thus defines an orientation i.e., a chirality. Thus, only a chiral system can have a non-zero $\kappa_C$.

To see why we expect a non-trivial topological response, e.g. $\kappa_C\neq 0$, we consider a very thin film with small curvature, i.e.  $K \xi^2 \ll 1$, where $\xi$ is the size of the Cooper pair. Then, the orbital spin of the Cooper pair (i.e., the spin of the pair due to orbital motion) is well defined and perpendicular to the surface. If the orbital spin of the pairs all have the same chirality the pair will respond to curvature in a similar way as to a magnetic field. In addition to the Aharanov-Bohm phase due to the charge $2e$ encircling the magnetic flux, the pair will also pick up the Berry phase $2\pi\chi l$, where $l$ is the orbital spin of the pair (we take $l > 0$ to denote right-handed rotation.) This means that the pair effectively responds to the combination of magnetic field and Gauss curvature, so that the Meissner effect will amount to expelling the combination $B+l K \Phi_0/4\pi$, rather than the magnetic field itself.

It is illuminating to see how the geometric Meissner effect emerges from a simple model, so we outline a derivation.
The spatial part of the Wen-Zee response can be obtained from the  
Ginzburg-Landau free energy for a vector order parameter $\varphi$ describing 
$p$-wave paired spinless fermions. \emph{Mutatis mutandis}, this model also applies to the 
spinfull case with half-vortices as described in Ref.~\cite{ivanov01}. The order parameter $\varphi$ can be written as,
\begin{align}
\varphi=\sqrt{\rho_+} e^{i\theta_+}(\hat e_{1}+i \hat e_{2})+\sqrt{\rho_{-}}e^{i\theta_{-}}( \hat e_{1}-i \hat e_{2})\ ,
\end{align}
where $\hat e_{1}$ and $\hat e_{2}$ are orthonormal basis vectors,  and $\rho_\pm$ and $\theta_\pm$ are densities
and phases of the two chiral components, respectively. We assume a Ginzburg-Landau free energy
\begin{align}
\!\!\!F=\int \! d^2x\sqrt g\biggl(\frac{ \hbar^{2}g^{ij}}{2m}(D_{i}\varphi)^{*}\!\cdot\! D_{j}\varphi+\!\frac{B^{2}}{2\mu_{0}}\!+\!V(|\varphi|)\biggr),\!
\end{align}
where $m$ is a mass, $\mu_0$ is the magnetic permeability, $B=1/\sqrt{g}\varepsilon^{ij}\partial_{i}A_{j}$ is the magnetic field scalar, $iD_{i}=(i\partial_{i}-2eA_{i}/\hbar)$, and roman indices denote spatial coordinates. We take a potential $V(|\varphi|)$, for which there is a mean-field solution with $\bar\rho=\bar\rho_+\neq 0$
and $\bar\rho_{-}=0$. Such a potential must exist for the flat geometry in order to at all have a $\chi$SC. By adiabatic continuity, such a solution will exist also for (at least weak) deformations of the surface, and since  $\kappa_{C}$ is quantized, it will remain fixed as long as there is no phase transition.

To lowest order we then get the London free energy,
\begin{align}
\! F_L=\int \! d^{2}x\sqrt{g} \Biggl(
\frac{4\hbar\bar\rho}{m}\left(\vec\nabla\theta_+ +\vec\omega-\frac{2e}\hbar \vec A\right)^2\!\!+\frac{B^{2}}{2\mu_{0}}\Biggr).\! \label{eq:londonfree}
\end{align}
The square of the vector within the parenthesis is determined by the metric $g$, and $\vec\omega$ comes from the derivatives of the basis vectors $\hat e_i$.
Varying $F_L$ gives  
\begin{align}
\left(\lambda_{L}^{2}\triangle-1\right)B & =\frac{\Phi_{0}}{4\pi}K\,,\label{eq:london}
\end{align}
where $\lambda_{L}=\sqrt{m/(32\bar{\rho}\mu_{0}e^2)}$ is the London length and $\triangle$ is the Laplace operator defined by the metric
$g$.
For a region with a linear size much larger than $\lambda_{L}$, 
we can average both sides to get Eq. \eqref{eq:kappa} with $\kappa_{C}=1$.
Chiral pairing in the $l$'th channel would give  $\kappa_{C}=l$. 
Note that this simple derivation does not give any flux-Hall response (see Ref. \onlinecite{moroz17}) term in the effective action. 

\begin{figure}
\includegraphics[width=\linewidth]{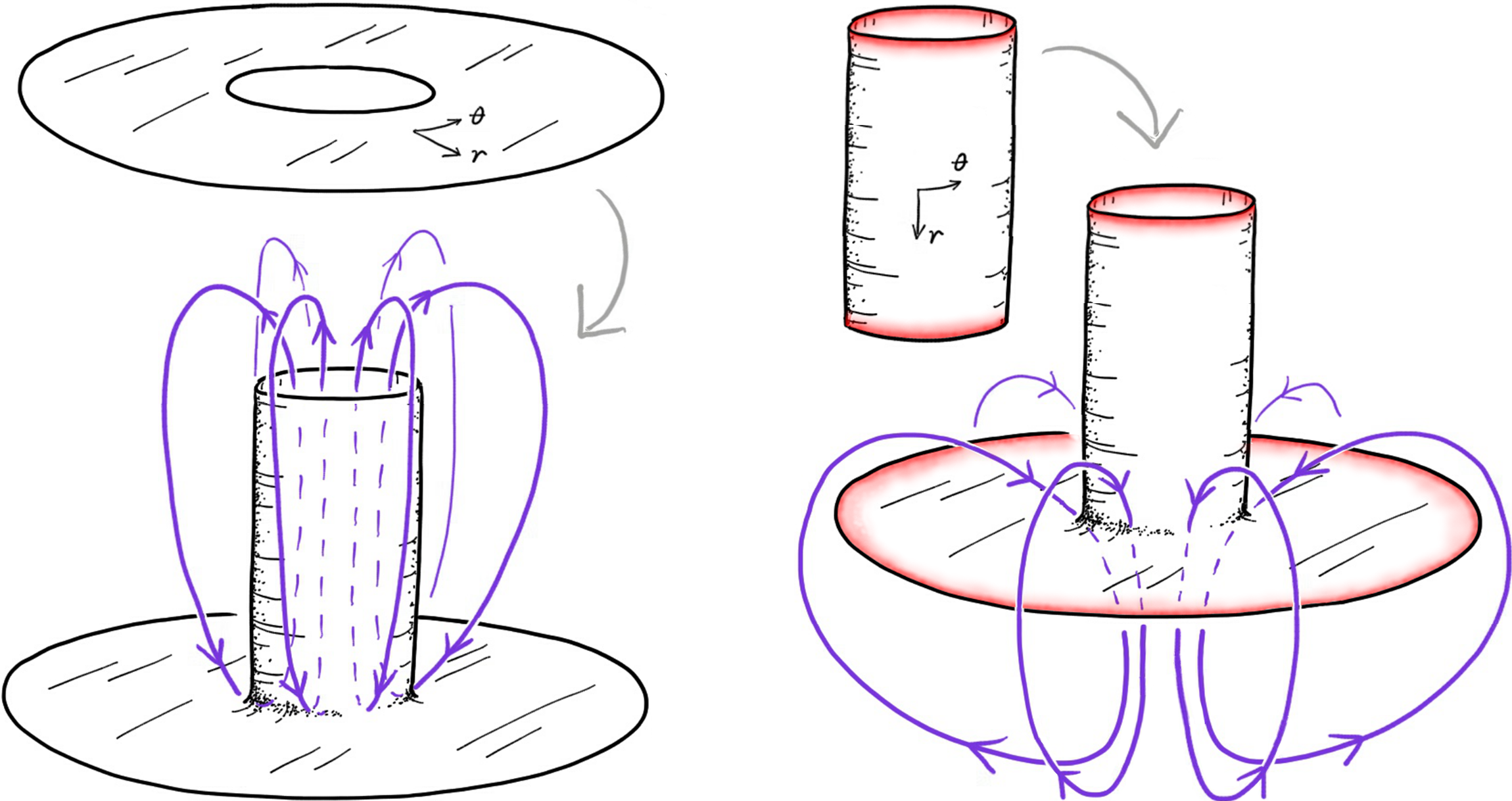} 
\caption{\textbf{Left.} Lifting the inner circle of an annulus to form a cylinder. \textbf{Right.}  Flattening the lower end of a cylinder to form an annulus. (Red indicates Majorino modes.)}\label{fig:disctocyl}
\end{figure}

\noindent \textbf{Thought experiments.}
\label{sec:though-exp}
Some apparently puzzling issues
 can be understood in terms of the geometric Meissner effect. 

\noindent \emph{1. Where are the Majorino edge modes?} 
A $\chi$SC generically has gapless edge modes in the thermodynamic limit, but \emph{odd pairing} $\chi$SCs can support zero-energy Majorana edge modes (\emph{Majorinos}), \emph{i.e.,}  an exact zero-energy mode for finite edge length (up to exponential corrections in system size).

For simple chiral $p$-wave SC models (such as in Ref.~\cite{read00}) the zero-flux state on the cylinder supports edge Majorinos and the state with a flux quantum through the cylinder has \emph{no} Majorinos. The opposite is true for the states on the annulus. 
Assuming an adiabatic change from the annulus to the cylinder, via
a tipless cone, the SC should remain in its ground state. That would mean
that Majorinos are either created or annihilated, which
would imply a closing of the bulk gap which, in turn, would contradict
the assumption of adiabaticity. In Ref. \cite{quelle16}, we resolved this
puzzle by showing that there is a level crossing and that the final state is not
the ground state. We now show that this is easily understood as a geometric Meissner effect.

Since the surface of a tipless cone is flat, one might think
that there would be no geometric Meissner effect. But $W_{WZ}$ in \eqref{geomeissner1}
depends on $\omega$, not on $K$, and the line integrals  $\int\omega_{i}dx^{i}$ are non trivial. With $\kappa_{C}=1$, the Wen-Zee term dictates that this geometric monodromy will be canceled by a flux 
through the hole of the cone. Going adiabatically from a cylinder  to 
a disc amounts to the spontaneous creation of a flux, and the 
edge Majorinos will remain. To stay in the ground state by an avoided crossing, would require the tunneling of a vortex across the SC, which is exponentially suppressed in the
system size. 

Alternatively, one can interpolate between
a cylinder and an annulus by gradually lifting the
inner edge of the annulus to form part of a cylinder and an associated region with total curvature $\int d^{2}x\:\sqrt{g}K=-2\pi$. When
the cylindrical region is longer than $\lambda_{L}$,  it follows from Eq.~\eqref{eq:london} that there will be a full  
flux quantum through the curved region. Since flux is conserved, it 
has to  enter somewhere and if the system is large, it must have
come from the inner edge, since tunneling from the outer edge is supressed. This is shown on the left side of Fig. \ref{fig:disctocyl}, where the surface is embedded in
$3d$ space and  the strength and sign of the $2d$ flux is illustrated by a  $3d$ field 
configuration (this configuration would be qualitatively correct for layered $3d$ films, but only if they are much thicker than $\lambda_L$). The flux lines are closing through the hole of the newly formed cylinder, 
such that the cylinder edge encircles
a  flux quantum while the annulus edge does not.
Neither of them support edge Majorinos, just as the annulus we started from!
If we do the opposite, \emph{i.e.,} flatten one end
of a cylinder, we end up on the state shown in the right side of Fig. \ref{fig:disctocyl},
that does support edge Majorinos.

\begin{figure}
\includegraphics[width=\linewidth]{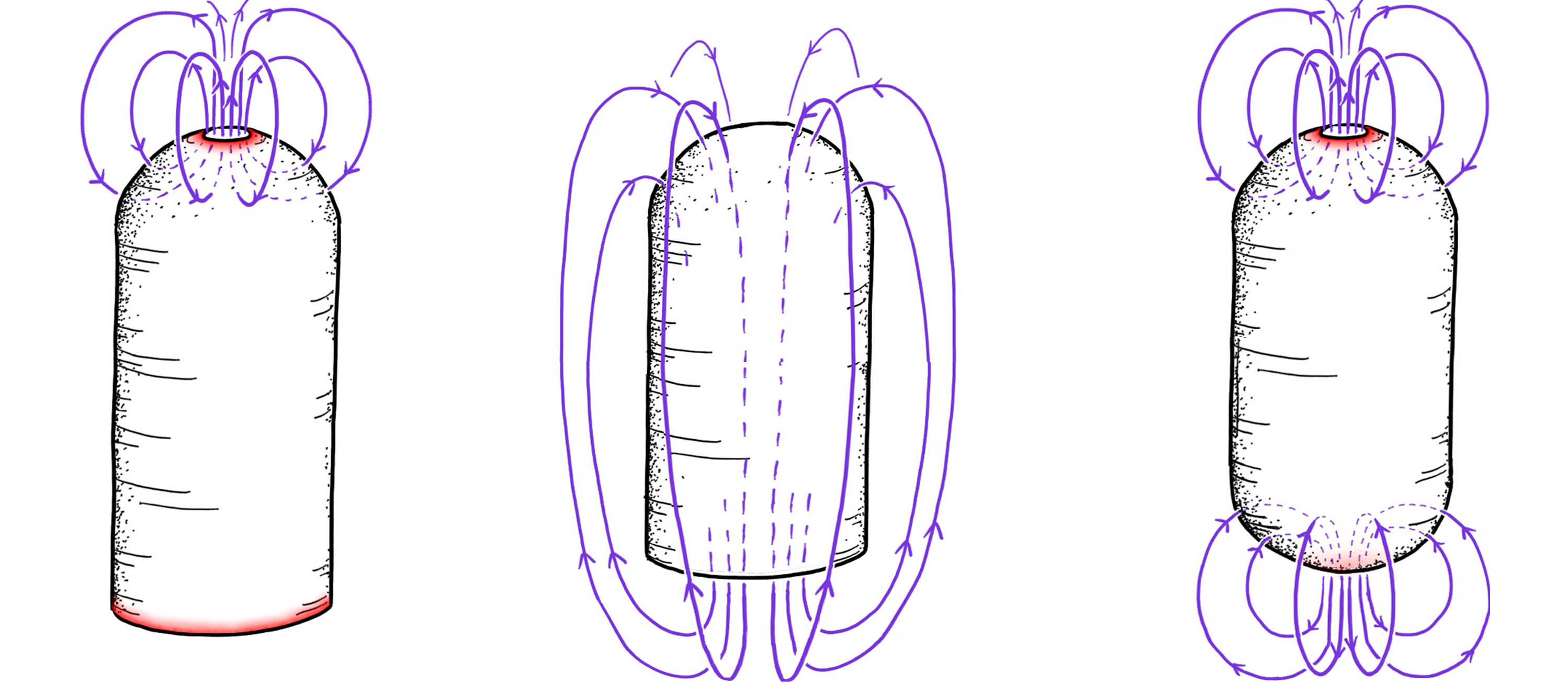} 
\caption{\textbf{Left.} Starting from a cylinder and closing one of the holes gives a magnetic flux at the cap. \textbf{Middle.} Starting from a half sphere and extending it, the flux points the opposite way, and extends all over the surface. This is not the ground state of the system due to the extended magnetic field lines. \textbf{Right.} As on the left, but closing both holes. (Red indicates Majorino modes.)} \label{fig:spheretocyl}
\end{figure}

Another perplexing question is what happens to the edge Majorino 
at the bottom of a cylinder, if the top is smoothly
capped. The resulting state has only one edge, so there should be no
Majorinos. But how can the bottom Majorino be removed by just a
local change at the other end? Again, we can understand what happens
by evoking the geometric Meissner effect. When we slowly deform an end of the
cylinder to a half sphere, we create curvature and thus flux. If
the cylinder is long enough, the flux must escape through the hole that
we are about to close. In the limit of a very small hole we do not get a
 homogeneous flux on the half sphere, but a vortex, and thus a \emph{localized 
Majorino}, as illustrated on the left side of Fig. \ref{fig:spheretocyl}. If we instead start from a sphere, and stretch one end out to form a cylinder, we
end up in the flux configuration shown in the middle of Fig. \ref{fig:spheretocyl}\textemdash a
state with no Majorinos. 

\noindent \emph{2. Closed manifolds.} If we close both ends of the cylinder, as on the right in Fig. \ref{fig:spheretocyl}, vortices arise. We can deform this geometry by shrinking the cylindrical section to zero to get a sphere. In this case, there is no symmetry to give preferred locations to the vortices, but since vortices in a type II superconductor repel, they would sit at antipodal points to minimize energy. Picking a direction of the line between them  amounts to a necessary spontaneous breaking of rotational symmetry.

That a \cscc on a sphere must have vorticity is an
effect analogous to the shift  in the relation \eqref{eq:qhshift} between flux and charge in  
 QH liquids.  
Here, it means that the number of flux-quanta through
a closed surface equals the integrated curvature $\chi$.

\begin{figure}
\begin{centering}
\includegraphics[width=\linewidth]{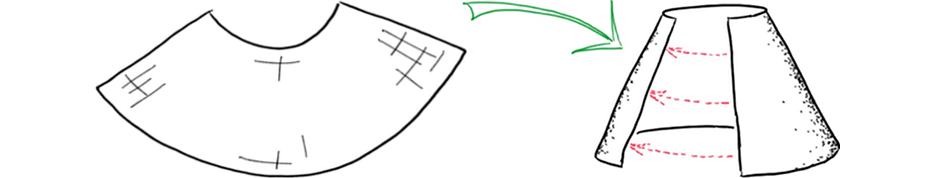}
\par\end{centering}
\caption{Folding a $\chi$SC, and creating a monodromy of the curvature form, results in a phase difference, and thus a current. The red arrows indicate the tunneling current.}
\label{fig:geojoseph} 
\end{figure}

\emph{3. The geometric Josephson effect.} Fig. \ref{fig:geojoseph} shows
how a cut cone can be formed by rolling up a segment cut from
a Corbino disc. Since the geometry is flat, the ground state of the
segment supports no flux, and neither does the partially rolled up
configuration shown to the right. But the cut cone,
obtained by gluing the disc along the dotted line, does support edge
currents. What happens is that
when the edges come close to each other, the system should be thought
of as a superconductor with a weak link that can maintain a phase
difference. As seen from Eq. \eqref{eq:londonfree}, $\omega$
enters just as an electromagnetic vector potential giving a geometric
version of the Josephson effect.

\emph{4. The geometric flux pump.} 
Laughlin notion of flux insertion in a Corbino geometry \cite{laughlin81} was historically
very  important for understanding  the integer quantization of the Hall conductance.
In the QH case, a unit electric charge is pumped from one edge to the other
by inserting a flux, and in the present context there is an analogous effect, which shows
that the Wen-Zee term must be quantized. 
The \emph{geometric flux pump} is operated by adiabatically transforming
a cylinder into an annulus and then back to the cylinder by pulling one of the edges through the other, which has the net effect of turning the cylinder inside out. In this process, the initial and final states have the same Hamiltonian, so if
it is adiabatic, the final state must be an energy eigenstate
below the bulk gap. This implies that there is an \emph{integer} number
of superconducting flux quanta through the hole of the cylinder, which
means that $\kappa_{C}$ must be quantized as an  integer.

\noindent \textbf{Experimental Realizations.}
Can the ggeometric Meissner effect be observed in the laboratory? Given a
candidate $\chi$SC, one can imagine several different experiments, depending
on the material to be tested. Interesting candidates are SrRuO$_4$ and bilayer graphene
intercalated with Ca, and to probe the symmetry of their order parameter one needs to grow them on a
concave or convex substrate. For bilayers, it is required that the substrate on which the graphene is
deposited must be non-superconducting, otherwise it will short-circuit the graphene, and destroy the geometric Meissner effect. Therefore, one could also conceive to suspend it on top of nano-pillars, as already experimentally realized. However, here it would be convenient
to have the nano-pillars forming a circular array of a radius $R$, instead
of a regular lattice, as in Ref.~\cite{otte15}. If the sample is much larger than the diameter of the circle, and can be
anchored outside, we expect a downward curvature in the inner region of the circle, simply
due to gravity. 
Another possibility would
be to set up a standing wave in a suspended sample and detect the
AC electromagnetic response. 

If the sample is a $2d$ sheet thinner than the London length $\lambda_{L}$, the 
screening of charges and fluxes changes from  exponential
to a power law at large distances. Since we still have Meissner-like
decay of the magnetic field, we would expect the relation \eqref{eq:london}
still to hold for regions with a radius $r \gg\lambda_L$, with corrections of order
$1/r$.

Comparing Eqs. \eqref{eq:qhshift} and \eqref{eq:kappa}, we see that in the QH case the 
geometric contribution to the flux is a small correction to the large dominant term due to the
background magnetic field, while in the \cscc case the geometric term stands alone.

The geometric Meissner effect scales proportional to the maximum bond-length stretching and inversely proportional to $\lambda^2_L$. With a maximum allowed bond-length stretching  of $1\%$ and $\lambda_L=1\mu$m, the magnetic field strength is of the order of $10\,\mu$T.
The best SQUIDs can detect fields as small as a pT; hence, such
a field should easily be detectable.

We hope that our work will motivate further experiments 
on curved $\chi$SC candidates, and contribute to the unveiling of this elusive
state of matter in an unequivocal and definitive way. 

\begin{acknowledgments}

We thank O. Golan, D. Mross and S. Moroz for helpful discussions and 
and S\"oren Holst for all the nice hand-drawn pictures.
\end{acknowledgments}

\bibliography{refs.bib}
\bibliographystyle{apsrev4-1}

\end{document}